\documentstyle[prl,aps,epsf,multicol]{revtex}
\begin{document}

\def\i{\imath\,}
\def\ih{\frac{\imath}{2}\,}
\def\undertext#1{\vtop{\hbox{#1}\kern 1pt \hrule}}
\def\ra{\rightarrow}
\def\lfa{\leftarrow}
\def\ua{\uparrow}
\def\da{\downarrow}
\def\Ra{\Rightarrow}
\def\lra{\longrightarrow}
\def\ler{\leftrightarrow}
\def\lrb#1{\left(#1\right)}
\def\O#1{O\left(#1\right)}
\def\EV#1{\left\langle#1\right\rangle}
\def\tr{\hbox{tr}\,}
\def\trb#1{\tr\lrb{#1}}
\def\dd#1{\frac{d}{d#1}}
\def\dbyd#1#2{\frac{d#1}{d#2}}
\def\pp#1{\frac{\partial}{\partial#1}}
\def\pbyp#1#2{\frac{\partial#1}{\partial#2}} 
\def\pd#1{\partial_{#1}}
\def\br{\\ \nonumber & &}
\def\brr{\right. \\ \nonumber & &\left.}
\def\inv#1{\frac{1}{#1}}
\def\be{\begin{equation}}
\def\ee{\end{equation}}
\def\bea{\begin{eqnarray}}
\def\eea{\end{eqnarray}}
\def\ct#1{\cite{#1}}
\def\rf#1{(\ref{#1})}
\def\EXP#1{\exp\left(#1\right)} 
\def\INT#1#2{\int_{#1}^{#2}} 
\def\LHS{left-hand side }
\def\RHS{right-hand side }
\def\COM#1#2{\left\lbrack #1\,,\,#2\right\rbrack}
\def\AC#1#2{\left\lbrace #1\,,\,#2\right\rbrace}

\def\bra#1{\left\langle#1\right|}
\def\ket#1{\left|#1\right\rangle}

\title{
Projected wave functions for fractionalized phases of quantum spin systems}
\author{D.~A.~Ivanov$^1$ and T.~Senthil$^2$}
\address{$^1$ Theoretische Physik, ETH-H\"onggerberg, CH-8093
Z\"urich, Switzerland \\
$^2$ Massachusetts Institute of Technology,
77 Massachusetts Ave.,  Cambridge 02139
}

\maketitle

\begin{abstract}
Gutzwiller projection allows a construction of 
an assortment of variational wave functions for 
strongly correlated systems. 
For quantum spin $S = 1/2$ models, Gutzwiller-projected wave functions have 
resonating-valence-bond structure and may represent states 
with fractional quantum numbers for the excitations. 
Using insights obtained from field-theoretical descriptions 
of fractionalization in two dimensions, 
we construct candidate wave functions of 
fractionalized states by projecting specific 
superconducting states. We explicitly demonstrate 
the presence of topological order in these states.   
 
\end{abstract}

\vspace{0.15cm}

\begin{multicols}{2}

\section{Introduction}
In the last several years, considerable theoretical 
progress has been made in understanding the 
possibility of fractional quantum numbers for the 
excitations of strongly interacting systems. A particular 
focus of attention has 
been on realizing such 
phases in quantum spin systems in two spatial dimensions\cite{PWA}. In conventional 
quantum phases of spin-$1/2$ magnets, such as 
an antiferromagnet or a spin-Peierls phase, 
the spin excitations carry quantum numbers $S = 1$ 
or higher multiples. In contrast, in a fractionalized phase, 
there are excitations (dubbed spinons) that carry spin $S = 1/2$. 
Various kinds of evidence for the presence of such fractionalized 
phases in spin models on a number of diverse lattices have been 
presented in Ref.~\onlinecite{frcspn}.
A detailed theoretical study\cite{RSSpN,Wen1,MudFr,NLII,z2long,frcmdl} 
of fractionalization in two dimensions has revealed that 
in the simplest cases such spinon excitations are necessarily 
accompanied by gapped ``vortex''-like excitations: visons. 
The visons carry no spin but have a long-range statistical 
interaction with the spinon: When a spinon is taken all the way 
around a vison, the wave function of the system changes sign. 
Thus the visons have an Ising-like character: 
two visons can annihilate each other and are equivalent to no vison at all. 
A concise mathematical description of this long ranged 
statistical interaction is given by assigning a 
$Z_2$ gauge charge to the spinons and a 
corresponding $Z_2$ gauge flux to the visons. 

A precise theoretical characterization\cite{Wen1,topth} 
of fractionalized phases serving to distinguish them 
from more conventional phases is provided by the notion 
of topological order --- a concept elucidated by Wen\cite{Wen2} 
in his work on the quantum Hall effect. 
When placed on a manifold with a non-trivial topology,
a fractionalized phase has a number of locally similar but globally distinct 
states that differ by whether or not a vison threads each hole of the 
manifold. 
 
If a spin model possesses a fractionalized phase, 
what does its wave function look like in terms of the original spins? 
In this paper, we 
argue that the wave functions describing a fractionalized state
may be obtained from the original RVB construction of Anderson\cite{PWA}.
We consider states obtained by 
projecting the wave function of a BCS superconductor (on a lattice)
to the Hilbert space of exactly one electron per site
(this procedure is also known as Gutzwiller projection\cite{Gutzwiller}).
Different topological sectors are obtained
by projecting states with/without superconducting vortices
(or, equivalently, with different boundary conditions). We 
motivate this construction from the 
field-theoretic description of fractionalized phases.
 
Projected wave functions have long been studied\cite{Yokoyama,Gros,Ivanov}
to gain insight into the properties of the cuprate materials.
However, as we show below, it is not guaranteed that 
the resulting wave function describes any fractionalized phase. 
The state that has been the subject of most attention is the 
``$d$-wave RVB state'' obtained by projecting a BCS $d_{x^2 - y^2}$ 
superconductor with nearest neighbor hopping and nearest neighbor 
pairing interactions. 
We argue below that such a superconducting state has some 
special symmetry properties due to which 
it is {\em not} expected to lead to a fractionalized state after 
projection (at half-filling). We confirm this by a direct 
numerical calculation on this state. 
It is not clear whether the nearest neighbor $d$-wave RVB state 
represents a good trial wave function for {\em any} stable phase of 
a spin system. 
Addition of a number of perturbations, such as for instance 
next-nearest-neighbor hopping, to the 
nearest-neighbor $d$-wave superconductor
removes the special symmetry properties --- the resulting state 
is then expected to lead, after projection, to a fractionalized state. 
We demonstrate explicitly the topological order in such projected states. 
A closely related state was recently considered in Ref.~\onlinecite{arun1}
at finite doping and was argued to provide a good 
description of cuprate phenomenology. We note that recent 
experiments\cite{Bonn,Wynn} place constraints on  the applicability of 
fractionalization ideas to the cuprates. 
Theoretical description of the cuprates with wave functions
corresponding to fractionalized states must hence be performed with
caution and examined for consistency with experiments.

Recently, a particular projected state has been argued to represent 
the ground state 
of the $J_1-J_2$ Heisenberg model on the square lattice\cite{Capriotti}. 
We argue that such a state is also expected to have topological order, 
and hence be fractionalized. We explicitly construct four 
globally distinct states on the torus to demonstrate the topological order.

If the projection does indeed lead to a fractionalized state, 
then we may construct wave functions for its excitations as follows. 
The distinct excitations of a superconductor are the BCS 
quasiparticles and the vortices.  
As originally suggested by Anderson\cite{PWA}, projection 
of the BCS quasiparticles directly leads to the spinons. 
We argue that the visons
may be obtained by projecting the wave function of the 
BCS $hc/2e$ vortex. This observation may be directly exploited to 
check for topological order in the projected wave functions. 
Consider putting the system on a torus. We may project the four
superconducting states obtained by threading or not threading a 
$hc/2e$ vortex through each hole of the torus. If the projected state
is fractionalized, and hence topologically ordered, 
these four states must be (after projection) orthogonal to each other 
in the thermodynamic limit though they are locally similar.

\section{Generalities on Gutzwiller projection} 
\label{ggp}

The original point of view on the Gutzwiller projected 
wave function\cite{PWA} is that it describes a 
state obtained by quantum disordering
a Heisenberg Neel antiferromagnet. However, as Gutzwiller projection 
(at half filling) freezes out the charge fluctuations inherent in the 
unprojected superconducting state, it may equally well be viewed as 
describing a state obtained by quantum disordering a superconductor. 
Recent theoretical work\cite{NLII,z2long} has sought to understand the 
properties of Mott insulators by viewing them as quantum disordered 
superconductors.
In this approach, insulating phases are regarded as condensates of the 
vortices of the superconductor. A fractionalized Mott insulator is 
obtained by pairing and condensing the BCS vortices while the unpaired 
$hc/2e$ vortex remains gapped. At energy scales well below the charge gap, the 
excitations in such a Mott insulator are the spinon and the vison. The 
spinons are precisely what become of the fermionic BCS
quasiparticles once the $hc/e$ vortices condense. The vison on the other 
hand is the remnant of the unpaired $hc/2e$ vortex 
- due to the $hc/e$ condensate, this survives with only a $Z_2$ character.

It is clear that this point of view 
fits in nicely with the idea of Gutzwiller 
projecting superconducting states 
to obtain wave functions for Mott insulators. 
However not all choices of the unprojected superconducting state are 
guaranteed to lead to fractionalized states on projection
(see Sections  below). In this section, we discuss the nature of 
projected vortex states and their use in constructing the 
wave function of the vison.

\subsection{RVB wave functions}

We begin by quickly reviewing the interpretation 
of the Gutzwiller-projected 
superconducting state as a resonating valence bond wave function. 
Consider the ground state of a BCS Hamiltonian on a
$L \times L$ square lattice ($L$ even):
\be
\label{hamiltonian}
H = -\sum_{ij} \sum_{\alpha} t_{ij} c^{\dagger}_{i\alpha}c_{j\alpha} +{\rm h.c.}
 + \Delta_{ij} c^{\dagger}_{i\ua} c^{\dagger}_{j\da} + {\rm h.c.}
\ee
Its wave function has the form
\be
\ket{BCS} = \prod_{\vec k} 
\left(u_{\vec k} + v_{\vec k} c^{\dagger}_{\vec k \ua} 
c^{\dagger}_{-\vec k \da}\right) \ket{0}
\label{BCS-wavefunction}
\ee
in standard notation. We restrict attention to spin-singlet, 
time-reversal-invariant ground states that also 
preserve all the symmetries of the lattice. The vectors $\vec k$ take 
values in the first Brillouin zone. We assume periodic 
boundary conditions in both spatial directions so that $\vec k =
\frac{2\pi}{L} (m_x , m_y)$ with $m_x, m_y$ integers 
(the system may be thought of as residing on a torus). 
We may rewrite this wave function as 
\be
\ket{BCS} \propto \exp\left( {\sum_k g(\vec k)  
c^{\dagger}_{\vec k \ua} c^{\dagger}_{-\vec k \da}} \right)\ket{0}
\ee
with $g(\vec k) = v_{\vec k}/u_{\vec k}$. 
So long as $g(\vec k) = g(-\vec k)$, this state is guaranteed to be a
spin singlet.

It will be useful to consider a first quantized version 
$\Psi(\{\vec x_i, \chi_i\})$ obtained by fixing the total 
number of particles to be $L^2$.
The $(\vec x_i, \chi_i)$ are the position and spin state of the 
$i$'th electron. We have 
\be
\Psi(\{\vec x_i, \chi_i\}) \propto \bra{\{\vec x_i, \chi_i\}} 
\left(\sum_k g_{\vec k}  c^{\dagger}_{\vec k \ua} c^{\dagger}_{-\vec k \da} 
\right)^{L^2/2} \ket{0}
\ee
The wave function $\Psi(\{\vec x_i, \chi_i \})$ is single-valued on the torus. 

The Gutzwiller-projected state $\ket{RVB}$ is obtained by 
projecting $\Psi$ further onto a subspace where there is 
exactly one particle per lattice site. The resulting wave function 
may be written as a sum over various valence-bond configurations 
on the lattice:
\be 
\ket{RVB} =P_G \ket{BCS} \propto \sum_{vbc} {\cal A}_{vbc} \ket{vbc}
\ee
where $\ket{vbc}$ denotes any particular valence bond covering. 
${\cal A}_{vbc}$ is the corresponding amplitude and is given by the product of 
$g(\vec x_i - \vec x_j)$ over all valence bonds connecting sites $(ij)$ 
appearing in the state $\ket{vbc}$. Here $g(\vec x_i - \vec x_j)$ is the 
Fourier transform of the function $g(\vec k)$ introduced above. 
 
\subsection{BCS vortices and their projections}

If the ground state of a spin system is correctly described by 
Gutzwiller projecting a BCS state, it is natural to construct 
excitations of the spin system by similarly projecting the 
excitations of the BCS superconductor. As argued by Anderson, 
projecting the BCS quasiparticle state leads to a wave function 
for neutral spin-$1/2$ spinons. The other distinct excitations of the 
superconductor are the vortices. What happens to these under the 
Gutzwiller projection? To answer this question, consider a wave function 
of a superconductor describing a state where a $nhc/2e$ vortex 
threads the torus along the $x$-direction. If $n$ is even, the 
(first-quantized) wave function may be taken simply as 
\be 
\Psi_{n_x} = \exp\left({\sum_{i} \frac{in \theta_i}{2}}\right)\,\Psi
\ee
where $\theta_i = \frac{2\pi y_i}{L}$ is the angular coordinate 
of the $i$'th particle. Thus, the vortex wave function 
for even $n$ may be simply obtained by multiplying the ground 
state wave function by a phase factor (which depends on the configuration of 
the particles).  
Upon Gutzwiller projection, there is one particle at every site; 
consequently the projected vortex state (for $n$ even) 
is {\em trivially} related to the projected ground state by an overall 
phase factor. Thus the projected even vortex state
does not lead to a distinct state of the spin system. 

For $n$ odd, the vortex wave function in the superconductor is not 
obtained by simply multiplying the ground state by a phase factor. 
Consider $n = 1$. The naive guess
\be
\exp\left({\sum_{i} \frac{i\theta_i}{2}}\right)\, \Psi
\ee
violates the physical requirement that all legitimate electron
wave functions must be single-valued on the torus. 
The correct wave function is constructed by multiplying by the phase 
factor above, {\em and} replacing $\Psi$ by a wave function that satisfies
antiperiodic boundary conditions along the $y$-direction (and periodic 
along the $x$-direction). Such a replacement may be obtained by 
considering the ground state of $H$ with antiperiodic boundary conditions 
along the $y$-direction:
\be
\ket{BCS^{+-}} = \prod_k \left(u_{\vec k'} + v_{\vec k'}c^{\dagger}_{\vec k' 
\ua} c^{\dagger}_{-\vec k' \da}\right) \ket{0} 
\ee
with $\vec k' = \vec k + \frac{\pi}{L}\hat{y}$ and 
$\vec k = \frac{2\pi}{L}(m_x, m_y)$ as before. 
Multiplying by the phase factor above has the effect of shifting 
all the momenta by $\frac{\pi}{L}$ along the $y$-direction.
A legitimate $hc/2e$ vortex state may therefore be constructed as
\be 
\ket{BCS'} = \prod_k \left (u_{\vec k'} + v_{\vec k'}c^{\dagger}_{\vec k 
+ \frac{2\pi \hat{y}}{L} \ua} c^{\dagger}_{-\vec k \da}\right) \ket{0} 
\ee
It is readily verified that this is a spin-singlet wave function. 
The first quantized $hc/2e$ vortex wave function (with $L^2$ particles)
may now be obtained straightforwardly as 
\be
\Psi' \propto \bra{ \{\vec x_i, \chi_i\} } \left( \sum_{\vec x, \vec x'} 
g'(\vec x, \vec x') c^{\dagger}_{\ua}(\vec x)c^{\dagger}_{\da}(\vec x') 
\right)^{L^2/2}\ket{0}
\ee
with 
\bea
g'(\vec x, \vec x')  & = & e^{i\frac{\pi}{L} (y + y')}g_{a.p}(\vec x-\vec x'),
 \\
g_{a.p}(\vec x -\vec x') & = & \sum_{\vec k'}g(\vec k')e^{i\vec k'. 
(\vec x - \vec x')}  \nonumber\\
& = & \sum_{\vec k'}g(\vec k') \cos\left(\vec k'. (\vec x - \vec x')\right)
\eea
Here, as before, $g(\vec k') = \frac{u_{\vec k'}}{v_{\vec k'}}$. 
The function $g_{a.p}(\vec x - \vec x')$ satisfies 
antiperiodic boundary conditions along the $y$ direction, and periodic 
along the $x$ direction:
\bea
g_{a.p}(\vec r + L\hat{y}) & = & -g_{a.p}(\vec r) \\
g_{a.p}(\vec r + L\hat{x}) & = & g_{a.p}(\vec r)
\eea
This vortex state may now be projected to obtain the new RVB wave function 
given by
\be
\ket{RVB'} = P_G \ket{BCS'} \propto \sum_{vbc} {\cal A}'_{vbc} \ket{vbc}
\ee
The amplitude ${\cal A}'_{vbc}$ is given by the product of 
$g'(\vec x_i - \vec x_j)$ over all valence bonds connecting 
sites $i$ and $j$ appearing in the 
state $\ket{vbc}$. This wave function can be further simplified 
by noting that the factor 
$\exp ({i\frac{\pi}{L}\sum_i y_i})$ arises in every term on the right-hand
side and is just a trivial multiplication by an overall phase factor, 
and hence may be dropped. 
(For an $L \times L$
square lattice with $L$ even, this phase factor is simply equal to one). 
This then amounts to 
Gutzwiller projecting the state $\ket{BCS^{+-}}$ obtained by changing 
the boundary conditions on the unprojected state. The resulting 
amplitude for a valence bond $(ij)$ is given by 
$g_{a.p}(\vec x_i - \vec x_j)$. 

Thus the projection of a $hc/2e$ vortex potentially survives as a 
non-trivial state in the spin system. In view of the discussion 
at the beginning of this section, it is clear that if the ground 
state described by $\ket{RVB}$ is indeed fractionalized, then $\ket{RVB'}$
describes a state with a vison threaded through along the $x$-direction. 
We will exploit this observation below to check for 
topological order in the projected wave functions.

\subsection{Short ranged RVB states and dimer models}

It is instructive to make a short digression, and specialize to 
superconducting states which are fully gapped, and where the function 
$g(\vec r)$ has a very short range of order a few lattice spacings.  
In that case, the resulting RVB wave function describes a state with only 
short ranged valence bonds (and presumably a full spin gap). 
For such a state, it is expected that the function $g(\vec k)$ is smooth 
in $\vec k$-space so that we may approximate
$g(\vec k') \approx g(\vec k)$. In real space, this amounts to
\be
g_{a.p}(\vec x - \vec x') \approx g(\vec x - \vec x') \cos\left(\frac{\pi}{L}
(y - y')\right)
\ee 
In the limit of large $L$, the cosine factor is one everywhere except 
for valence bonds that connect sites to the left of 
$y = L$ to sites to the right of $y =L$ for which it is $-1$. Thus the 
difference between the projected ground state and the 
projected $hc/2e$ state may be summarized as follows: Consider drawing 
a vertical line parallel to the $x$-direction that cuts all the links 
of the lattice between $y = L$ and $y = 1$. In the projected vortex state, 
the amplitude for any valence bond that 
cuts this line acquires a factor $(-1)$ relative to the ground state. 
In other words, the amplitude for any valence bond configuration 
where an odd number of valence bonds intersect this line is $(-1)$ relative 
to the ground state. 
We note that this construction of the projected vortex wave function reduces 
to the state considered by 
Read and Chakraborty in their pioneering early work\cite{ReCh} on topological 
order in RVB wave functions. 
Unfortunately, as already recognized in that paper\cite{ReCh} it is not clear 
that the particular RVB state 
discussed there describes any stable phase of a spin system. 

An approximate description of a short ranged RVB state is through a quantum 
dimer model\cite{qdm}. 
It is well-known\cite{qdm,Bstl} that the quantum dimer model on a torus has 
four distinct topological sectors classified by whether the number of dimers 
intersecting
particular lines similar to the one introduced above is even or odd. 
Fractionalization in the original spin model
is signalled by topological order in the quantum dimer model when the four 
topological sectors become 
degenerate in energy.  
In this case, these even/odd sectors clearly correspond to the 
symmetric/antisymmetric combination 
of the state with no vison and a threaded vison. This is completely 
consistent with our construction of the vison wave function by projecting 
the $hc/2e$ vortex. 

Our primary interest in this paper will be long-ranged RVB states where the 
spinons are gapless. We will explore the properties of  
wave functions for such states by numerical calculations using the 
construction above.

\section{Topological order: general considerations}
\label{to}

Though the projected $hc/2e$ vortex survives as a non-trivial 
state of the spin system, it is still not necessary 
that the state $\ket{RVB}$ has topological order.
The presence or lack thereof of topological order may be established by 
examining the following two conditions:

{\bf (i)} $\langle RVB'|RVB\rangle = 0$ 
as the system size goes to infinity, and 

{\bf (ii)} The expectation values of all local physical 
operators be the same in both states $\ket{RVB}$ and $\ket{RVB'}$ 
again in the thermodynamic limit. 

The first condition is closely related to the presence of a vison gap in the bulk. 
Indeed, the states $|RVB\rangle$ and $|RVB'\rangle$ differ
by one vison tunneled across the cylinder (or torus), and therefore
$\langle RVB|RVB'\rangle$ may be interpreted as the
amplitude of such a tunneling event.
The second condition guarantees that the distinction between the states is 
not in any local properties but rather in global ones. 
In this section, we will examine some simple arguments that motivate 
the choice of particular superconducting wave functions which will
lead to fractionalized states on projection. 

Consider the first condition. It is easy to check that the overlap
$\langle BCS'|BCS \rangle$ 
goes to zero very rapidly as $L$ becomes large. This is 
expected physically 
as the vorticity is a good quantum number for the unprojected BCS state. 
However, this does {\em not}
guarantee $\langle RVB'|RVB \rangle = 0$ due to the projection. 
To get better insight 
into what is needed for condition (i) to be satisfied, 
we will employ the following useful characterization of the Gutzwiller 
projection. In the unprojected Hilbert space, the 
states at each site are $\ket{0}, \ket{\ua}, \ket{\da}, \ket{\ua\da}$ 
in obvious notation. 
Projection keeps only the states $\ket{\ua}, \ket{\da}$
at each site. At each site $i$, introduce the physical spin operator 
$\vec S_i $ and the pseudo-spin operator $\vec T_i$:
\bea
\vec S_i & = & \frac{1}{2} c^{\dagger}_i \vec \sigma c_i \\
T^z_i & = & \frac{c^{\dagger}_i c_i - 1}{2} \\
T^{+}_i & = & c^{\dagger}_{i\ua}c^{\dagger}_{i\da} \\
T^{-}_i & = & c_{i\da}c_{i\ua}
\eea
Here $T^{\pm}_i = T^x_i \pm T^y_i$. The $\vec T_i$ satisfy $SU(2)$ 
commutation relations just like the physical spin operators $\vec S_i$. 
Furthermore, all components of $\vec T_i$ commute with all components of 
$\vec S_i$.  Clearly the states $\ket{\ua}, \ket{\da}$ are singlets under the 
pseudo-spin rotation generated by the $\vec T_i$, while the states 
$\ket{0}, \ket{\ua\da}$ form a doublet under the pseudospin rotation. 
Thus the Gutzwiller projection is equivalent to projection onto the 
singlet sector (at each site) of the pseudospin $SU(2)$ rotation.
This has the general implication
\be
P_G \ket{\psi_{\rm unproj}} =P_G U\ket{\psi_{\rm unproj}}
\ee
for an arbitrary $SU(2)$ rotation
\be
U = e^{i\sum_i \vec \theta_i . \vec T_i}
\label{SU2-rotation}
\ee
where the parameters $\vec \theta_i$ may be chosen independently
on each site. 
Thus a local pseudospin $SU(2)$ rotation of the unprojected state does not 
change the state after projection.
We will therefore call this a gauge rotation.

In view of the above, a sufficient condition 
for the orthogonality of $\ket{RVB}$ and $\ket{RVB'}$ is simply
\be
\label{orcrt1}
\langle BCS|U|BCS' \rangle = 0
\ee
for any arbitrary pseudospin $SU(2)$ rotation (\ref{SU2-rotation}). 
(We note that this is not  in general a {\em necessary} condition 
for the orthogonality).  

The discussion has so far been general.
Eqn.~(\ref{orcrt1}) imposes some conditions on the nature of the BCS state 
which could lead to a wave function
for a fractionalized state after projection. However Eqn.~(\ref{orcrt1}) 
is still not in a form which is directly useful 
in providing guidance in writing down such states. To get a more useful form, 
we specialize to a particular class of 
unprojected states. In general, the state $\ket{BCS}$ will be a linear 
superposition of 
states with different total particle number. On the other hand, the 
projected state $\ket{RVB}$ 
has exactly one particle per site, and hence an exact total of $L^2$ 
particles on a lattice with $L^2$ sites. 
Now, for a general BCS state, if we plot the probability distribution 
of having a total of $n$ particles, it will have a 
sharp peak at some average value $n_0$ and will die rapidly 
for $|n - n_0|$ large. If $n_0$ is different from $L^2$, then the 
Gutzwiller projection picks out the tails of the original BCS wave function. 
In this case, the projected wave function may have very little to do with
the unprojected one. For the projected state to retain the significant 
features of the spin physics of the 
unprojected state, it is clearly advantageous\cite{thank} to require that the 
mean number of particles (before projection) is $L^2$. If this is 
satisfied, the projection will have a gentler effect than if the mean 
number is different from one per site. 
To further soften the effect of the projection, it is clearly 
advantageous to require that the mean number be one for {\em every} 
unprojected state obtained from $\ket{BCS}$ by a pseudospin $SU(2)$ 
gauge rotation. This ensures that the projection does not pick up the 
tails of the wave function in any gauge. 

The requirement that the average electron number on each site is one
may be simply written as
\be
\langle T_i^z \rangle = 0
\ee
on all sites $i$. Further the requirement that 
this be true for every gauge-rotated ground state 
is equivalent to requiring that 
\be 
\label{mfcstr}
\langle \vec T_i \rangle =0,
\ee
i.e.\ the average values of the generators of the gauge rotations
vanish.

From now on, we will specialize to unprojected states where the requirement 
Eqn.~(\ref{mfcstr}) is satisfied. In this case, 
we may hope that we can use our intuition about the unprojected state to 
infer the properties of the projected state. When do we expect that 
Eqn.~(\ref{orcrt1}) will be satisfied for such a BCS state? 
Note that $\ket{BCS'}$ 
differs from $\ket{BCS}$ in having a $hc/2e$ vortex threaded 
through the hole of the cylinder. If the ground state of a physical system 
is given by the superconducting state $\ket{BCS}$, we
can label it's excitations by their total vorticity quantum number. 
The state $\ket{BCS'}$ has total vorticity $\pm 1$ compared to 
$\ket{BCS}$, and hence is orthogonal. However, so long as 
$U\ket{BCS}$ is also a superconducting state, 
it will also have a fixed vorticity 
which differs from that of  $\ket{BCS'}$ by an odd number. 
Consequently, in this case Eqn.~(\ref{orcrt1}) will be satisfied. 
If however for some $U$, the gauge rotated state $U\ket{BCS}$ 
is not a superconducting state, then it may be regarded as a 
coherent superposition of electron wave functions 
each of which carry a definite vorticity. (The vorticity is not a 
good quantum number in a non-superconducting state). It is then 
possible that $\ket{BCS'}$ is not orthogonal to that particular 
gauge rotated state, so that Eqn.~(\ref{orcrt1}) is not 
satisfied. At the very least, we lose the general reason 
requiring orthogonality. 

Based on the reasoning above, we make the following conjecture:

\medskip

{\em An unprojected state where the constraint} (\ref{mfcstr}) {\em is 
satisfied, and which cannot be rotated to a non-superconducting
state by an $SU(2)$ 
gauge rotation, will lead to a topologically ordered state 
after projection}.

\medskip
 
This conjecture provides strong motivation for choosing particular 
superconducting states that we may project to obtain topologically ordered
states. On the other hand, we do not provide any formal proof of this 
conjecture in this paper. It is supported by the reasoning in this 
section, and  by our numerical results. 

Here we need to make a clear distinction between ``superconducting''
and ``non-superconducting'' states. Formally, we call a state
``gauge-equivalent to non-superconducting'' 
if it is invariant under a global $U(1)$
subgroup of the full group of gauge rotations (\ref{SU2-rotation}).
Indeed, a non-superconducting state is characterized by a fixed number
of electrons. Then the $U(1)$ group of {\em global} rotations by $T^z$
(i.e., the subgroup generated by the sets $\vec\theta_i=\theta\hat z$)
produces only trivial multiplications by phase factors leaving the
state invariant. We call such a state a ``$U(1)$ state'' [or even a
``$SU(2)$ state'' in the case when the maximal subgroup leaving the
unprojected state invariant is $SU(2)$]. In this terminology, a
state that is superconducting in any gauge is called a ``$Z_2$ state''.
This classification is a sub-classification of a more detailed one
introduced by Wen\cite{Wen3}.

A set of conditions for the $U(1)$ invariance (being
gauge-equivalent to non-superconducting) 
of a given state may be formally written
in terms of projector operators onto occupied quasiparticle states.
Consider the set of Bogoliubov--deGennes doublets 
$(u_{\vec k}, v_{\vec k})$
participating in the ground state (\ref{BCS-wavefunction}). We can
define a projector onto the linear subspace spanned by those doublets.
In real space, this projector may be thought of as a set of $2\times 2$
matrices $P_{ij}$ labeled by the site indices $i$ and $j$ and defined as
\be
P_{ij} = \sum_{\vec k} \pmatrix{u_{\vec k}(i) \cr v_{\vec k}(i)}
                       \pmatrix{u_{\vec k}^*(j) & v_{\vec k}^*(j)} 
\ee
Under the $SU(2)$ gauge transformation, these matrices transform
as  
\be
P_{ij} \ra U_i P_{ij}U^{\dagger}_j
\ee
For a non-superconducting state, all matrices $P_{ij}$ are simultaneously diagonal.
To see when a given state can be gauge rotated to such a non-superconducting state,
it is useful to consider the chain products of such
matrices starting and ending at the same point $i$:
\be
\label{defac}
A_i[C] = \Pi_C \left(P_{ij} P_{jk} \dots P_{li} \right)
\ee
where $(i,j,k,\dots,l,i)$ define a closed curve on the lattice. 
$A_i[C]$ is a $2 \times 2$ matrix for each 
lattice site $i$ and for each closed curve $C$ starting and ending at that
site. Now one easily verifies that all matrices $P_{ij}$ may
be simultaneously diagonalized if and only if for any $i$ and
for any pair of closed contours $C$ and $C'$ both starting and ending at
$i$,
\be
\label{orcrt2}
\left[A_i[C], A_i[C'] \right] = 0
\ee
Thus to check that a given wave function describes a $Z_2$ state
(i.e. cannot be gauge rotated to a non-superconducting form),
it is sufficient to verify that $A_i[C]$ and  $A_i[C']$ do not commute
for at least one choice of $i$, $C$, and $C'$.

Instead of checking whether the {\em wave function} may be rotated
to a non-superconducting form, one may perform a similar test
for the {\em Hamiltonian} (\ref{hamiltonian}). The Hamiltonian can
be rotated to a non-superconducting one ({\em i.e} containing no pairing terms) if and only if
\be
\left[B_i[C], B_i[C'] \right] = 0,
\label{B-condition}
\ee
for all $i$ and $C$, where
\be
B_i[C] = \Pi_C \left(H_{ij} H_{jk} \dots H_{li} \right),
\quad
H_{ij}=\pmatrix{t_{ij} & \Delta_{ij} \cr \Delta^*_{ij} & -t^*_{ij}}
\label{Hij}
\ee
This is a convenient {\em sufficient} condition for being
a $U(1)$ wave function, but not a necessary one: in certain
cases, a superconducting Hamiltonian may have a non-superconducting
ground state (with a definite particle number).

It is instructive to  consider some specific examples of 
superconducting states to see how these conditions work in practice. 
Consider for instance a nearest 
neighbor $d$-wave superconductor where $t_{ij}, \Delta_{ij}$ are 
non-zero only on nearest-neighbor bonds, and take the values 
\bea
t_{ij} & = & t \\
\Delta_{ij} & = & \pm\Delta .
\eea
The plus sign in the second equation is 
for horizontal bonds, and the minus sign for vertical bonds. 
It is readily seen that the condition in 
Eqn.~(\ref{mfcstr}) is satisfied by this state. However, as is also easily seen,
all the $B_i[C]$ commute for this state. This is of 
course consistent with the well-known fact that this 
state can be gauge rotated to a pure hopping state 
(known as the staggered flux state). 
We therefore expect that this state will not lead 
to a fractionalized state after projection. This is 
confirmed by our numerical calculations below. 

Now consider adding the next-near-neighbor diagonal hopping 
$t'$ to this state. The resulting Hamiltonian still describes a
$d_{x^2 - y^2}$ superconductor. However, a simple 
calculation shows that there are two non-commuting $B_i[C]$ matrices 
so that this can no longer be rotated to pure hopping. 
Addition of $t'$ changes the mean density away from one particle per site. 
This can however be compensated by adding an on-site chemical 
potential term to the unprojected Hamiltonian. The state thus 
constructed is therefore a good candidate for projecting to 
get a wave function for a fractionalized state. Below, in Section \ref{num} 
we demonstrate this by a numerical calculation. 

Surprisingly, the criteria obtained above for the superconducting 
state to describe a topologically ordered state after projection
may also be motivated from a completely different point of 
view\cite{Wen1,Wen3}. Consider any spin $S = 1/2$ model in two dimensions. 
As is well-known, 
it is possible to use a representation of the spins in terms of 
fermionic spin-$1/2$
operators. This representation is exact so long as the constraint that 
the fermion occupation is one at each site
is imposed on the Hilbert space. A popular approach is to treat 
the resulting fermion Hamiltonian 
in mean field theory. At the mean-field level, the excitations are 
neutral spin-$1/2$ fermions 
described by a Hamiltonian of the general form Eqn.~(\ref{hamiltonian}). 
The exact constraint inherent in the fermionic representation is replaced 
precisely by Eqn.~(\ref{mfcstr}). This mean field state is capable of 
correctly describing a possible physical phase of the spin model as 
long as it is stable to fluctuations.
As discussed by Wen\cite{Wen1,Wen3}, the criteria for the 
stability of the mean field state to fluctuations are precisely 
that expressed in Eqn.~(\ref{B-condition}). If the mean field state 
is indeed stable to fluctuations, then we expect that the 
candidate wave function for the physical state described by it is given 
by the Gutzwiller projection of the mean field
wave function to the physical Hilbert space.

\section{Numerical results}
\label{num}

We further verify our conjecture  numerically 
by testing the conditions (i) and (ii) of Section \ref{to} 
on the square lattice in
the toroidal geometry for several examples of the projected BCS wave
functions. We label the ground states in the four topological sectors
as $\left|{+}{+}\right\rangle$, 
$\left|{+}{-}\right\rangle$, $\left|{-}{+}\right\rangle$, and
$\left|{-}{-}\right\rangle$, according to the boundary conditions
imposed in $x$ and $y$ directions.
We employ the variational Monte Carlo method described in detail in   
Ref.~\onlinecite{Gros} applied to the square $L{\times}L$ tori. 

\begin{figure}
\epsfxsize=0.9\hsize
\centerline{\epsfbox{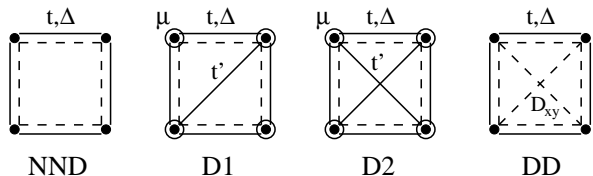}}
\medskip
\narrowtext
\caption{The mean-field states generating the four wave functions
NND, D1, D2, and DD. The terms of the Hamiltonian (\ref{hamiltonian})
are shown on one plaquette of the lattice. The solid lines denote
hopping, the dashed lines denote pairing, and the circles around
vertices denote the chemical potential.
}
\label{fig-four-wavefunctions}
\end{figure}

We consider four different types of projected wave functions: 
the nearest-neighbor $d_{x^2-y^2}$ state and its three modifications
by including hopping or pairing along the plaquette
diagonals (Fig.~\ref{fig-four-wavefunctions}). All these
wave functions may be parameterized by translationally 
invariant Hamiltonians
(\ref{hamiltonian}), and we can conveniently describe them in terms
of the Fourier transform of $H_{ij}$ (defined above in Eq.~(\ref{Hij})):
\begin{equation}
H({\bf k})=\pmatrix{\xi({\bf k}) & \Delta({\bf k}) \cr
\Delta^*({\bf k}) & -\xi(-{\bf k}) }\, .
\end{equation}

The nearest-neighbor $d_{x^2-y^2}$-wave BCS state (further denoted
as NND) is defined by its kinetic and pairing amplitudes:
\begin{eqnarray}
\xi({\bf k}) &=& -2(\cos k_x + \cos k_y)\, , 
\label{NND-xi} \\
\Delta({\bf k}) &=& \Delta_0 (\cos k_x - \cos k_y)\, .
\label{NND-Delta}
\end{eqnarray}
In the projected NND wave function, the nearest-neighbor antiferromagnetic
correlations are maximized at the intermediate value of 
$\Delta_0 \approx 0.55$ \cite{Yokoyama,Dmitriev} (the optimal values of
$\Delta_0$ reported in these two references differ by
several percent; the precise value of $\Delta_0$ is not important
for our qualitative results concerning topological order).
We find that the NND state after projection has
no topological order. This agrees with our conjecture, since the
NND state can be rotated to the pure-hopping staggered-flux state
by a SU(2) gauge rotation \cite{Zhang,Wen-Lee}.

\begin{figure}
\epsfxsize=\hsize
\centerline{\epsfbox{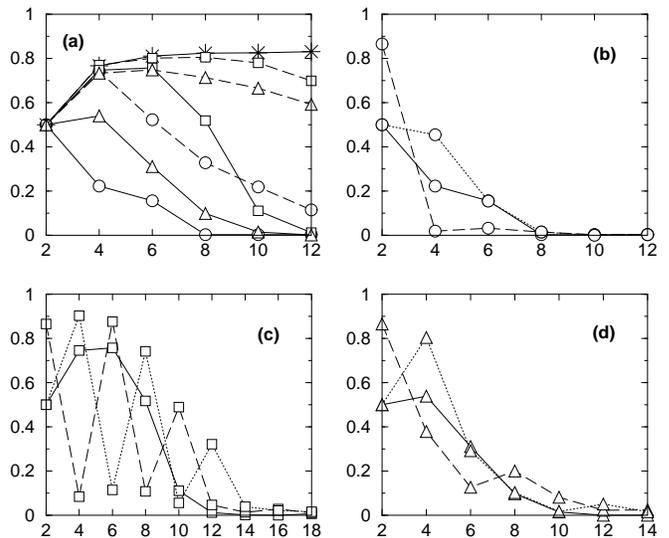}}
\medskip
\narrowtext
\caption{Overlaps between wave functions with different boundary conditions
on tori $L\times L$ as a function of the system size $L$. All wave functions
have $\Delta_0=0.55$.
{\bf (a)}: overlaps $\langle {+-} | {-+} \rangle$. 
Stars: NND wave function.
Circles: D1 wave functions (dashed line: $t'=0.5$, $\mu=0.417$;
solid line: $t'=1.0$, $\mu=0.696$). Squares: D2 wave functions
(dashed line: $t'=0.3$, $\mu=0.509$; solid line: $t'=0.5$, $\mu=0.794$).
Triangles: DD wave functions (dashed line: $D_{xy}=0.4$;
solid line: $D_{xy}=1.0$).
{\bf (b)}: overlaps for the D1 wave function with  
$t'=1.0$, $\mu=0.696$.
{\bf (c)}: overlaps for the D2 wave function with  
$t'=0.5$, $\mu=0.794$.
{\bf (d)}: overlaps for the DD wave function with  
$D_{xy}=1.0$. 
In plots {\bf (b)}--{\bf (d)},
solid line corresponds to $\langle {+-} | {-+} \rangle$,
dashed line --- to  $\langle {+-} | {++} \rangle$,
dotted line --- to $\langle {+-} | {--} \rangle$.
For these three wave functions, the overlap  
$\langle {++} | {--} \rangle$ is found
to be zero for all system sizes, within the numerical
accuracy (about 0.003).
}
\label{fig-overlap}
\end{figure}

In addition to the NND state,  we consider its three modifications which
are $Z_2$ states (cannot be gauge rotated to a non-superconducting
form). In the first modification (we denote it
by D1), we add hopping
along one of the plaquette diagonals which amounts to replacing (\ref{NND-xi})
by
\begin{equation}
\xi({\bf k}) = -2[\cos k_x + \cos k_y+t'\cos(k_x+k_y)]-\mu \, .
\end{equation}
The chemical potential $\mu$ is added for adjusting the average particle
density in the unprojected wave function in order to satisfy
the mean-field constraint (\ref{mfcstr}).

The second wave function (dubbed D2) is analogous to the previous
one, but with hopping along both plaquette diagonals:
\begin{eqnarray}
\xi({\bf k}) &=& -2[\cos k_x + \cos k_y \nonumber \\
&& +t'(\cos(k_x+k_y)+\cos(k_x-k_y))]-\mu \, .
\end{eqnarray}

The third wave function is $d_{x^2-y^2}+d_{xy}$ wave function
proposed by Capriotti et al \cite{Capriotti} as a variational ansatz
for the $J_1$--$J_2$ Heisenberg model. This wave function
(denoted further as DD)
has the nearest-neighbor form (\ref{NND-xi}) of $\xi({\bf k})$,
together with $\Delta({\bf k})$ involving both nearest-neighbor
pairing and pairing along the plaquette diagonals:
\begin{equation}
\Delta({\bf k}) = \Delta_0 (\cos k_x - \cos k_y+2 D_{xy} \sin k_x \sin k_y)\, .
\end{equation}
One easily verifies that the projected DD wave function has all the
symmetries of the square lattice, even though the unprojected wave
function does not. This wave function obeys the mean-field constraint
(\ref{mfcstr}) without a chemical potential term.

\begin{figure}
\epsfxsize=\hsize
\centerline{\epsfbox{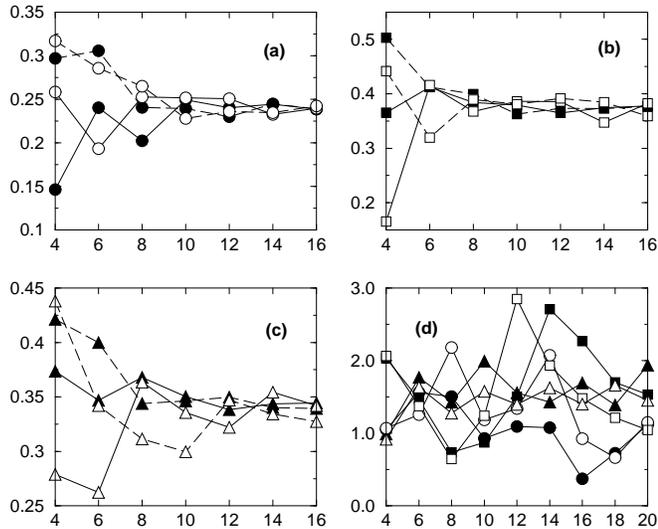}}
\medskip
\narrowtext
\caption{
Nearest-neighbor spin correlations $S_{\alpha\beta}$ for 
wave functions with different boundary conditions
on tori $L\times L$ as a function of the system size $L$. 
All wave functions have $\Delta_0=0.55$.
{\bf (a)}: D1 wave function with 
$t'=1.0$, $\mu=0.696$.
{\bf (b)}: D2 wave function with
$t'=0.5$, $\mu=0.794$.
{\bf (c)}: DD wave function with
$D_{xy}=1.0$.
In plots {\bf (a)}--{\bf (c)}:
solid symbols, solid lines correspond to $S_{+-}$;
solid symbols, dashed lines --- to $S_{-+}$;
empty symbols, solid lines  --- to $S_{++}$;
empty symbols, dashed lines --- to $S_{--}$.
{\bf (d)}: the spread of local correlations $\Delta S$ 
as a function of the system size $L$.
Solid circles: D1 wave function, $t'=1.0$, $\mu=0.696$.
Empty circles: D1 wave function, $t'=1.2$, $\mu=0.744$.
Solid squares: D2 wave function, $t'=0.5$, $\mu=0.794$.
Empty squares: D2 wave function, $t'=0.6$, $\mu=0.909$.
Solid triangles: DD wave function, $D_{xy}=1.0$.
Empty triangles: DD wave function, $D_{xy}=1.2$.
In all plots, the error bars are smaller than the symbol
size, except for the rightmost points in plot (d) where
the error bars are of the order of the symbol size.
}
\label{fig-spin}
\end{figure}

The first test for the topological order is the overlap
of the two wave functions on a torus projected from the mean-field
states with different boundary conditions.
Note that when nodes
in the mean-field excitation spectrum fall on points of the
momentum lattice, the wave function becomes ill-defined.
For our choice of lattice on the tori, this happens for
the NND state where the nodes are fixed
at ($\pm{\pi\over2}$,$\pm{\pi\over2}$) points. Therefore, not all four
topological sectors are realized for the NND state (for our
choice of the lattice placement).
However, the sectors $+-$ and $-+$ are well-defined for any $L\times L$
torus, and we take the overlap between these two wave functions
as the first check of the topological order. Alternatively,
this overlap may be viewed as the overlap of the wave function with
its reflection with respect to the $x$--$y$ diagonal.

We have calculated this overlap with the use of the
variational Monte Carlo procedure\cite{Gros}.
The results for the NND wave function
and for its three modifications D1, D2, and DD,
are presented in Fig.~\ref{fig-overlap}a.
In all cases, we take $\Delta_0=0.55$;
for D1 wave functions we take $t'=0.5$ ($\mu=0.417$) and
$t'=1.0$ ($\mu=0.696$); for D2 wave functions we take $t'=0.3$
($\mu=0.509$) and $t'=0.5$ ($\mu=0.794$); for DD wave functions
we take $D_{xy}=0.4$ and $D_{xy}=1.0$.
We see that the overlap for the NND wave
function saturates at large system size (which implies
the absence of topological order), while for the three
other wave functions it decreases to zero with increasing system
size. This is the first signature of the topological order in those
wave functions. As a side remark, we mention that if we force
$\mu=0$ in the D2 wave function, the projected wave function
has no topological order (the overlap saturates at large $L$),
but such a wave function violates
the mean-field-density constraint (\ref{mfcstr}), and
we do not discuss it here.

For the wave functions deep in the ``topologically ordered'' phase,
we can also verify the orthogonality of all the four topological
sectors $++$, $+-$, $-+$, and $--$. The four corresponding overlaps are
plotted for each of the wave functions D1, D2, and DD in
Fig.~\ref{fig-overlap}b--d.
All the overlaps for these wave functions decrease to zero
with increasing system size, which indicates orthogonality of the
four sectors.

The second test for the topologically ordered state is the
equality of local expectation values between different
topological sectors (condition (ii) of Section~\ref{to}). 
We verify this condition for the nearest-neighbor spin-spin correlations.
In the four topological sectors on the torus, there are
four different nearest-neighbor correlators: 
\begin{eqnarray}
S_{+-} &\equiv& -\langle {+-}| \sigma^z_i \sigma^z_{i+\hat{\bf x}} 
|{+-}\rangle, \nonumber \\ 
S_{-+} &\equiv& -\langle {+-}| \sigma^z_i \sigma^z_{i+\hat{\bf y}} 
|{+-}\rangle, \nonumber \\
S_{++} &\equiv& -\langle {++}| \sigma^z_i \sigma^z_j |{++}\rangle, \\
S_{--} &\equiv& -\langle {--}| \sigma^z_i \sigma^z_j |{--}\rangle. \nonumber
\end{eqnarray}
In a topologically ordered state, the difference between
these four quantities must rapidly decay with increasing system size.
We compute these correlators numerically for the same 
three wave functions used before in the overlap computations.
The results are presented in Fig.~\ref{fig-spin}a--c
demonstrating that the four correlations indeed converge to 
a single value in large systems. This supports the contention that the distinction between 
the states is in global properties and not in local ones. 

To quantify the rate of this convergence, we consider the
mean square deviation of the four quantities $S_{+-}$, $S_{-+}$, $S_{++}$, 
and $S_{--}$. In Fig.~\ref{fig-spin}d, 
we plot this mean square deviation multiplied
by the number of lattice sites
\begin{equation}
\Delta S \equiv L^2 \left[ {\sum_{\alpha\beta} S_{\alpha\beta}^2 \over 4}
- \left( {\sum_{\alpha\beta} S_{\alpha\beta} \over 4} \right)^2 \right]^{1/2}
\end{equation}
as a function of system size.
The finite-size effects are very strong because of the
nodes in the spectrum. To clarify the size dependence of $\Delta S$,
we take one more wave function in each of the three classes D1, D2, DD,
with slightly different variational parameters
(in addition to the wave functions considered previously).
The data in Fig.~\ref{fig-spin}d 
indicate that $\Delta S$ remains approximately
independent of system size. This corresponds to the difference in local
correlations $S_{\alpha\beta}$ decaying as $L^{-2}$ with
system size.

This slow convergence of correlations among different
topological sectors can be explained already at the mean-field
level from the nodal singularities in the spectrum. At the
nodal points, the vector $(u_{\vec k}, v_{\vec k})$ has
a strong singularity: as the wave vector $\vec k$ goes around the
nodal point, the vector $(u,v)$ rotates by half turn in the
plane. This singularity translates into the $L^{-2}$ dependence
of local correlations on the boundary conditions (changing the
boundary conditions amounts to shifting the lattice of vectors
$\vec k$). From our numerical results we see that this
rate of convergence is preserved by the projection. This may serve
as an indication that in the $Z_2$ states the projection preserves
the nodal spinons, as suggested by Wen\cite{Wen3}. This $L^{-2}$
law applies to the general case of local correlations. The
energy expectation value is a very special local correlation which
has a weaker singularity at the nodal point. At the mean-field
level, the energy splitting between different topological sectors
can be found to be $L^{-3}$ (per site). This translates into the splitting
in {\em total} energy decaying as $L^{-1}$. We expect that this
asymptotic law also holds after the projection. To verify this
numerically, one needs to compute the expectation values of the
actual ``energy'', which is a correlation function {\em optimized}
for a particular Hamiltonian by adjusting the variational parameters. 
In this paper we do not identify spin Hamiltonians for which
our wave functions are optimal, and thus are not able to verify
our expectation for the energy splitting.
 Note that the $L^{-1}$
convergence in energy also means convergence of the energy
of an individual vison, i.e., deconfinement of visons which is
necessary for the topological order.

We have also tested all the above classes of wave functions for the
valence-bond crystal (``spin-Peierls'') ordering\cite{Fradkin}.
We have computed the correlations of the $z$-component of the
singlet order parameter $\langle S_z(i) S_z(i+{\bf x}) S_z(j) S_z(j+{\bf x})
\rangle$ in systems as large as 18$\times$18. We find this correlation
function rapidly decaying with increasing the distance $|i-j|$.
The decay is slowest for the D2 wave function for which it appears to be
close to $R^{-2}$ (for the $t'=0.5$, $\mu=0.794$ wave function) or 
faster (for the $t'=0.6$, $\mu=0.909$ wave function). For D1 and DD
wave functions the decay of such correlations is much stronger than in the
D2 wave function, which indicates the absence of the
valence-bond ordering. Note that the D2 wave functions for the values
of parameters considered in this paper exhibit a relatively large
correlation length for the overlaps between topological sectors
(see Fig.~\ref{fig-overlap}c for the data on the $t'=0.5$, $\mu=0.794$
wave function). Therefore one may expect that correlation functions
have different behavior at larger and smaller length scale, and
that our computations in relatively small systems are
therefore not completely reliable for determining 
the correct long-distance behavior of the correlations. 
A more detailed analysis
of the valence-bond crystal ordering and of its interplay with
the topological order is left for future study.

\section{Conclusion}

In this paper, we have formulated conditions for the presence of topological
order in RVB systems and have verified them for several specific
examples of Gutzwiller-projected wave functions. Our results suggest
that appropriate Gutzwiller-projected wave functions may represent ground
states of fractionalized phases of spin systems.

This work is only the first step towards describing the topologically
ordered RVB states. For a better understanding of the properties
of the topological order, a more extensive quantitative study is needed.
It should include an analysis of correlation lengths involved in the
conditions (i) and (ii) for the topological order (in particular,
on cylinders/tori with different aspect ratios). Variational
wave functions may provide a useful tool for studying quantum
phase transitions between states with and without topological order.
An extremely interesting question in this respect is the possibility
of the coexistence of the topological order simultaneously with
the antiferromagnetic Neel order or with the valence-bond crystal
order. Such a coexistence of a topological and a conventional ordering
may be tested by projecting 
appropriate superconducting states with the spin or translational
symmetry broken before projection.

Of course, the study of the variational wave functions have physical
implications only when the Hamiltonians are identified for which
those wave functions are good trial states. Our test for the
topological order may provide a guidance in the
search for microscopic spin Hamiltonians that exhibit fractionalized
ground states.

\bigskip

We thank  P.~A.~Lee, X.-G.~Wen, A.~Ioselevich, M.~Feigelman,
L.~Ioffe, G.~Blatter, M.~P.~A.~Fisher, and O.~Motrunich 
for useful discussions. We are grateful 
to A.~Paramekanti, M.~Randeria, and N.~Trivedi for sharing their unpublished results 
and for several useful conversations. This
work was supported by MRSEC program of 
the National Science Foundation under grant DMR-9808941 and by the Swiss National Foundation.

\end{multicols}

\begin{references}

\bibitem{PWA} P.~W.~Anderson, Science {\bf 235}, 1196 (1987).

\bibitem{frcspn}
R.~Coldea, D.~A.~Tennant, A.~M.~Tsvelik, and Z.~Tylczynski,  
Phys. Rev. Lett. {\bf 86}, 1335 (2001); 
G.~Misguich {\em et al}, Phys. Rev. B {\bf 60}, 1064 (1999); 
W.~LiMing {\em et al}, Phys. Rev. B {\bf 62}, 6372 (2000);
S. Sachdev, Phys. Rev. {\bf B45}, 12377 (1992);
C.~H.~Chung, J.~B.~Marston, and R.~H.~McKenzie, J. Phys. C {\bf 13}, 
5159 (2001); 
C.~H.~Chung, J.~B.~Marston, and S.~Sachdev, Phys. Rev. B {\bf 64}, 
134407 (2001); R.~Moessner and S.~L.~Sondhi, Phys. Rev. Lett. {\bf 86}, 1881 (2001);
L.~Balents, M.~P.~A.~Fisher, and S.~M.~Girvin, cond-mat/0110005.

\bibitem{RSSpN} 
N.~Read and S.~Sachdev, Phys. Rev. Lett. {\bf 66}, 1773 (1991);
S.~Sachdev and N.~Read, Int. J. Mod. Phys. B {\bf 5}, 219 (1991).

\bibitem{Wen1} 
X.~G.~Wen, Phys. Rev. B {\bf 44}, 2664 (1991).

\bibitem{MudFr}C. Mudry and E. Fradkin, Phys. Rev. {\bf B49}, 5200 (1994).

\bibitem{NLII} 
L.~Balents, M.~P.~A.~Fisher, and C.~Nayak, 
Phys. Rev. B {\bf 60}, 1654 (1999); 
{\em ibid.} {\bf 61}, 6307 (2000).

\bibitem{z2long}
T.~Senthil and M.~P.~A.~Fisher,  Phys. Rev. B {\bf 62}, 7850 (2000).

\bibitem{frcmdl}
T.~Senthil and O.~Motrunich, cond-mat/0201320.

\bibitem{topth} 
T.~Senthil and M.~P.~A.~Fisher, Phys. Rev. B {\bf 63}, 134521 (2001).

\bibitem{Wen2}
X.~G.~Wen, Int. J. Mod. Phys. B {\bf 4}, 239 (1990).

\bibitem{Gutzwiller}
M. C. Gutzwiller, Phys. Rev. Lett. {\bf 10}, 159 (1963);
Phys. Rev. A {\bf 134}, 1726 (1965).

\bibitem{Yokoyama}
H.~Yokoyama and H.~Shiba, J.~Phys.~Soc.~Jpn. {\bf 57}, 2482 (1988);
H. Yokoyama and M. Ogata, J. Phys. Soc. Jpn. {\bf 65}, 3615 (1996).

\bibitem{Gros}
C. Gros, Phys.~Rev.~B 38 (1988), 931; Ann. Phys. {\bf 189}, 53 (1989).

\bibitem{Ivanov}
D. A. Ivanov, P. A. Lee, and X.-G. Wen,
Phys. Rev. Lett. {\bf 84}, 3958 (2000).

\bibitem{arun1}
A. Paramekanti et. al., Phys. Rev. Lett., {\bf 87}, 217002 (2001).

\bibitem{Bonn} D.~A.~Bonn {\em et.al}, Nature {\bf 414}, 887 (2001). 

\bibitem{Wynn} J.~C.~Wynn {\em et.al}, Phys. Rev. Lett. {\bf 87}, 197002 (2001).

\bibitem{Capriotti}
L. Capriotti, F. Becca, A. Parola, and S. Sorella,
cond-mat/0107204.

\bibitem{ReCh} N. Read and B. Chakraborty, Phys. Rev. {\bf B40}, 7133 (1989).
See also Ref.~\onlinecite{Bstl}. 

\bibitem{qdm} 
D.~S.~Rokhsar and S.~A.~Kivelson, Phys. Rev. lett. {\bf 61}, 2376 (1988).

\bibitem{Bstl} N. Bonesteel, Phys. Rev. {\bf B40}, 8954 (1989).

\bibitem{thank}
We are grateful to X.G. Wen for emphasizing this to us. 

\bibitem{Wen3}
X.~G.~Wen, cond-mat/0107071.

\bibitem{Dmitriev}
D. V. Dmitriev, V. Ya. Krivnov, V. N. Likhachev, and A. A. Ovchinnikov,
Fiz. Tv. Tela. {\bf 38}, 397 (1996). [Phys. Solid State {\bf 38}, 219 (1996)]

\bibitem{Zhang}
F. C. Zhang, C. Gros, T. M. Rice, and H. Shiba,
Supercond. Sci. Technol. {\bf 1}, 36 (1988).

\bibitem{Wen-Lee}
X.-G. Wen and P. A. Lee, Phys. Rev. Lett. {\bf 76}, 503 (1996).

\bibitem{Fradkin}
E.~Fradkin,
{\it Field Theories of Condensed Matter Systems}
(Addison--Wesley, Redwood City, CA, 1991).


\end{references}
\end{document}